\documentclass[runningheads]{llncs}
\usepackage[T1]{fontenc}
\usepackage{graphicx}
\usepackage[hidelinks,colorlinks=true,linkcolor=blue,citecolor=blue,urlcolor=blue]{hyperref}
\usepackage{url}
\usepackage{float}
\usepackage{booktabs} 
\usepackage{amsmath}

\begin{document}
\title{GPU Kernel Optimization Beyond Full Builds: An LLM Framework with Minimal Executable Programs}
\titlerunning{An LLM Framework with Minimal Executable Programs}
%
\author{\anonymous}
\institute{\anonymous}
\author{Ruifan Chu \and Anbang Wang \and Xiuxiu Bai \and Shuai Liu \and Xiaoshe Dong}

\authorrunning{R. Chu et al.}
%
\institute{School of Software Engineering, Xi'an Jiaotong University, China
\email{refun@stu.xjtu.edu.cn}}
\maketitle              

\begin{abstract}
In high-performance computing, hotspot GPU kernels are primary bottlenecks, and expert manual tuning is costly and hard to port. Large language model methods often assume kernels can be compiled and executed cheaply, which fails in large applications where full builds and runs are expensive. We present an end-to-end LLM framework with performance feedback that optimizes kernels without building the full application. From independently extracted hotspot kernels, it automatically completes code into a Minimal Executable Program (MEP), then performs multi-round iterative optimization and evaluation outside the full application. The framework integrates Automatic Error Repair and Performance Pattern Inheritance to fix faults, preserve correctness, reuse effective tiling/memory/synchronization strategies, and reduce search cost. Optimized variants are reintegrated into the original application for validation. We evaluate on NVIDIA GPUs and the Haiguang Deep Computing Unit (DCU) platform (AMD-licensed architecture) using PolyBench, the AMD APP SDK, and hotspot kernels from large-scale supercomputing applications. The method achieves average speedups of 5.05$\times$ (PolyBench on NVIDIA), 7.77$\times$ (PolyBench on DCU), 1.77$\times$ (AMD APP SDK), and 1.25$\times$ on three hotspot kernels, surpassing direct LLM optimization. The approach requires no full-source dependencies, offers cross-platform portability, and enables practical, low-cost GPU kernel optimization.
\end{abstract}


\section{Introduction}
In high-performance computing (HPC), particularly in heterogeneous parallel computing, hotspot kernels are often the primary source of performance bottlenecks~\cite{ref1,ref2,ref3}. Whether in scientific simulation, numerical computation, or large language model (LLM) inference and training, the performance of hotspot kernels directly determines overall system efficiency~\cite{ref4,ref5}. However, GPU kernel optimization has long relied on expert experience and manual optimization, which is time-consuming and challenging; when hardware architectures change, substantial rewrites are often required~\cite{ref1,ref2,ref3,ref4,ref5,ref6}. Although compiler-based automation has somewhat lowered the barrier to entry, it still cannot fully replace expert knowledge and often requires costly parameter tuning to approach peak hardware performance~\cite{ref7,ref8,ref9}.

A key prerequisite for effective kernel optimization is code completion for isolated kernels. Existing work on code completion largely targets general-purpose programming and IDE assistance. However, specialized completion for HPC and GPU kernels often fails to adequately capture low-level hardware characteristics, parallel execution models, and specific dependency environments~\cite{ref10}. Moreover, in complex engineering settings or with non-standard interfaces, LLM-based completion still faces challenges in ensuring correctness and executability, such as missing dependencies, data type mismatches, improper thread configurations, and insufficient boundary handling~\cite{ref11}. These issues not only prevent successful compilation and execution but also introduce uncertainty into downstream performance optimization. In high-cost HPC contexts, relying solely on LLMs for code completion often fails to preserve functional semantics and meet the stability demands of performance optimization.

Meanwhile, LLM capabilities in code completion, software automation, and program optimization are rapidly advancing~\cite{ref12,ref13}, demonstrating strong potential for GPU kernel optimization~\cite{ref10,ref14,ref15,ref16,ref17}. Nevertheless, a critical limitation of most LLM optimization studies is their assumption of structurally complete, directly runnable kernels, which enables repeated compile-run cycles for performance feedback~\cite{ref10,ref14,ref15,ref16,ref17}. This assumption often breaks down in large production applications, where building, linking, and running the full application can be extremely expensive, with a single run taking minutes to hours. Under such high costs, validating every LLM-generated candidate within the full application is unsustainable in terms of development time and computational resources~\cite{ref8,ref18}.


To address the aforementioned limitations, particularly the high cost of validation in production environments, we propose an iterative optimization framework with performance feedback that enables automated kernel optimization without requiring full application builds. Our method integrates four key components:

\begin{enumerate}
\item \textbf{Minimal Executable Program (MEP)} to isolate the hotspot kernel under time and data-size constraints for repeatable evaluation.
\item \textbf{Performance-Feedback Iterative Optimization} to generate multi-version candidates, profile them, and rank via trimmed-mean measurements each round.
\item \textbf{Diagnostics-Guided Automatic Error Repair (AER) and Functional Equivalence (FE)} to fix build/runtime faults and enforce correctness throughout the loop.
\item \textbf{Performance Pattern Inheritance} to retain effective tiling/memory/synchronization strategies and accelerate convergence across rounds and platforms.
\end{enumerate}

\begin{figure}[t]
    \centering
    \includegraphics[width=\textwidth]{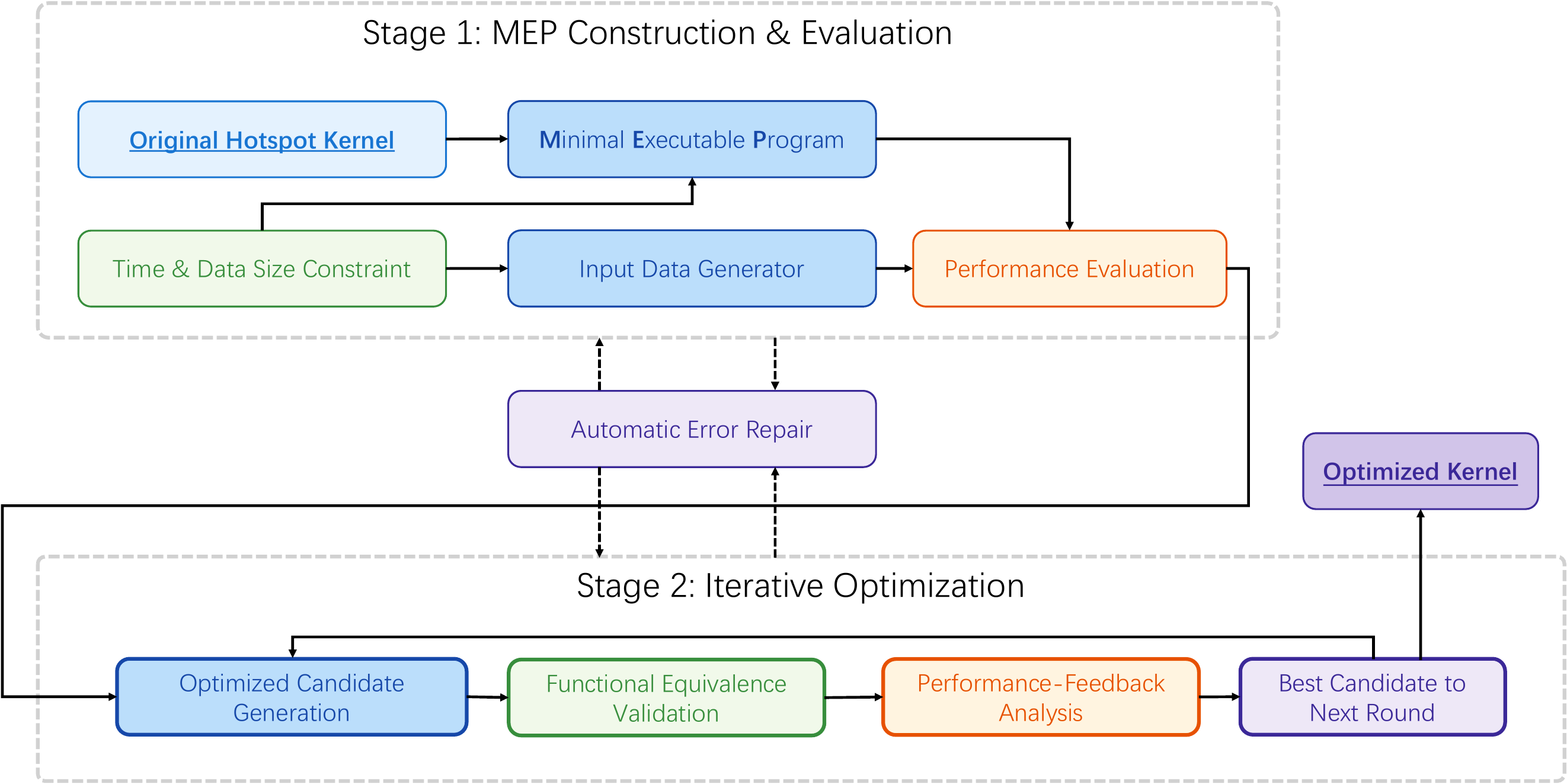}
    \caption{The MEP-based LLM framework for GPU kernel optimization.}
    \label{fig:overall}
\end{figure}

This design enables efficient kernel-level optimization within a standalone MEP environment, achieving robust performance improvements while preserving correctness. The overall framework is illustrated in Figure~\ref{fig:overall}.

\section{Related Work}

\subsection{Compiler-Based and Learning-Based Approaches for GPU Kernel Optimization}
Compiler frameworks and domain-specific languages have long been primary tools for GPU kernel optimization. Systems such as Halide~\cite{ref19}, TVM~\cite{ref8}, MLIR~\cite{ref20}, TensorFlow XLA~\cite{ref21}, and NVIDIA CUTLASS~\cite{ref22} provide high-level abstractions for expressing tensor computations and use compiler techniques to optimize code performance.

Several frameworks further incorporate auto-optimization and search strategies, such as AutoTVM~\cite{ref8} and Ansor~\cite{ref9}, enabling automatic exploration of optimization configurations with minimal human intervention. More recent systems, including Triton~\cite{ref17} and Mirage~\cite{ref7}, leverage new intermediate representations and multi-level optimization techniques to achieve performance that is competitive with expert-level implementations.

Despite these advances, compiler-based methods still face challenges in cross-platform generalization, scalability, and optimization cost. Transferring or re-optimizing kernels across different architectures often requires significant manual effort and architecture-specific rewriting~\cite{ref3}.

\subsection{LLM-Based Approaches for GPU Kernel Optimization}
With the rapid development of LLMs, researchers have increasingly explored their potential for automating GPU kernel optimization. Early studies primarily focused on general-purpose programming tasks, while subsequent work has expanded to performance-sensitive code completion, including loop vectorization~\cite{ref23}, assembly-level optimization~\cite{ref24}, and tensor program optimization~\cite{ref25}.

In the domain of GPU kernel completion~\cite{ref10,ref17}, iterative optimization has emerged as a dominant paradigm: the LLM produces candidate kernels, which are then refined through compilation checks, correctness verification, and performance profiling. A central challenge in this process is maintaining functional equivalence~\cite{ref26} while achieving performance improvements comparable to those of hand-optimized implementations.

Recent efforts have investigated prompt-driven optimization strategies~\cite{ref27} and reinforcement learning-based frameworks~\cite{ref28}. While these methods achieve promising results on small-scale benchmarks, they typically assume that compilation and execution can be performed at low cost. In large-scale HPC environments, where building and running the full application are computationally expensive~\cite{ref8,ref18}, such assumptions limit their practicality and efficiency.

\section{Methodology}

\subsection{Minimal Executable Program Construction and Evaluation}

In real HPC or AI applications, hotspot GPU kernels are often embedded in large projects where performance data collection depends on full builds and runs, both incurring high costs. We therefore automatically complete independently extracted hotspot kernels into a MEP, eliminating dependence on the original full application.

During code completion, if compilation errors, runtime errors, or output inconsistencies occur, the system feeds the erroneous code, diagnostics, and abnormal outputs back to the LLM to trigger AER. The same AER mechanism also applies during the iterative optimization stage.

\begin{figure}[t]
    \centering
    \includegraphics[width=0.9\textwidth]{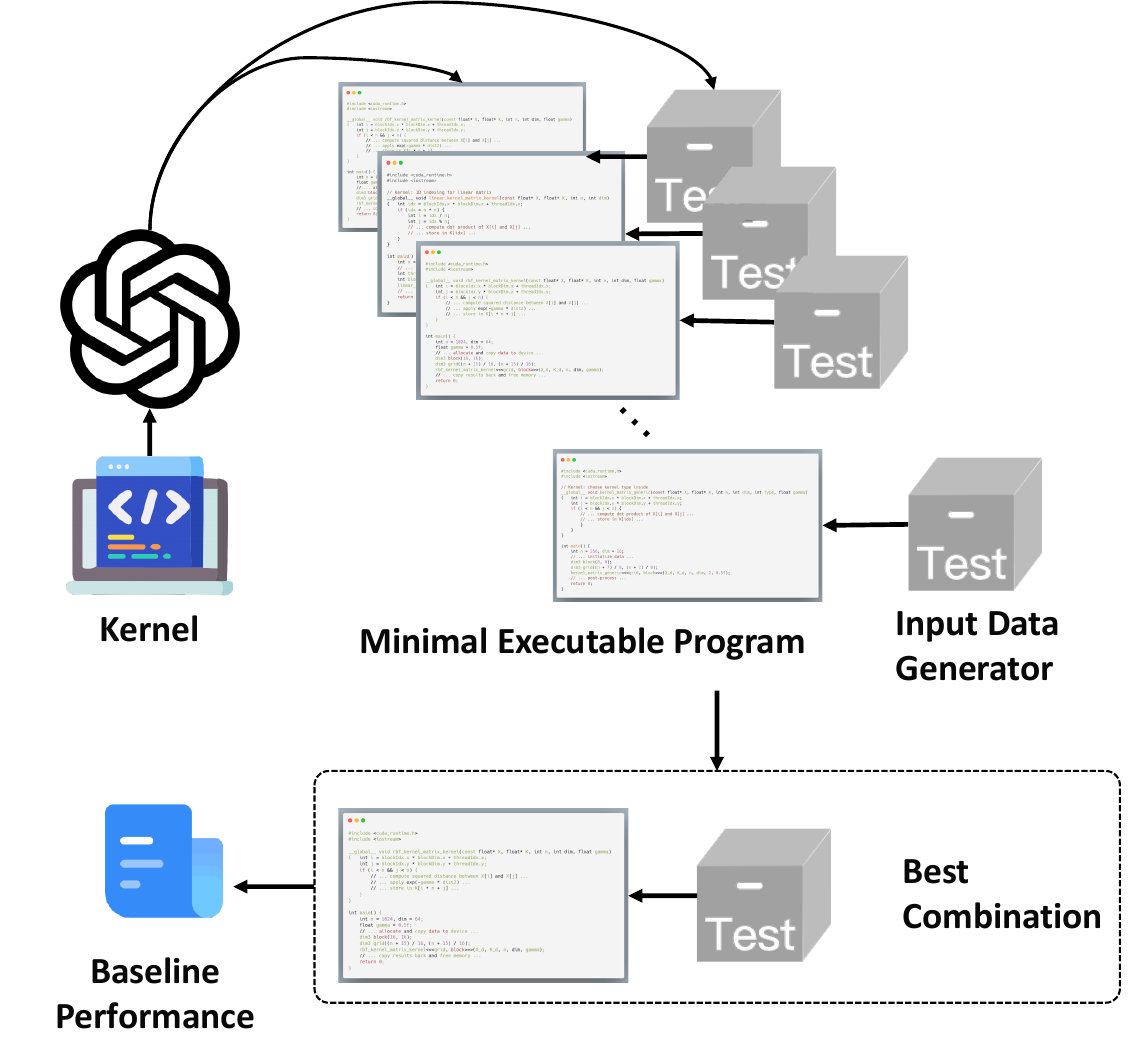}
    \caption{The framework of MEP construction and evaluation.}
    \label{fig:stage1}
\end{figure}

The framework of MEP construction and evaluation is illustrated in Figure~\ref{fig:stage1}.

\subsubsection{Minimal Executable Program Construction}

This step builds a MEP for independently extracted hotspot kernels. The MEP preserves functional logic and introduces sufficient workload for stable performance analysis, decoupling evaluation from the large application.

To ensure measurability and efficiency for subsequent optimization, the completion must satisfy two execution-time constraints:
\begin{equation}
\left\{
\begin{aligned}
T_{\text{ker}} &\ge T_{\text{min}},\\
T_{\text{overall}} &\le T_{\text{max}}.
\end{aligned}
\right.
\end{equation}
where $T_{\text{ker}}$ denotes the average kernel execution time of the MEP, $T_{\text{min}}$ the minimum significance threshold, $T_{\text{overall}}$ the total MEP execution time, and $T_{\text{max}}$ the global upper bound. The first condition ensures that kernel execution is long enough to reveal meaningful performance differences. The second ensures that the total MEP execution time remains within an acceptable budget.

The completion process is LLM-driven: after parsing kernel parameters and logic, the framework assembles a runnable program that handles input preparation, memory allocation, kernel invocation, and result handling; it also selects a problem size that satisfies the stated time constraints.

\subsubsection{Input Data Generator Construction}

Once the MEP is generated, we construct an Input Data Generator that matches the kernel's input pattern, enabling repeatable and accurate performance evaluation in the MEP environment.

The generator must satisfy the data size constraint
\begin{equation}
S_{\text{data}} \le S_{\text{max}},
\end{equation}
so that the total execution time remains within limits and memory use is controlled. Here, $S_{\text{data}}$ is the size of the generated input data (e.g., in bytes or MB), measured and reported, and $S_{\text{max}}$ enforces compliance with the $T_{\text{overall}} \le T_{\text{max}}$ constraint specified in Section~3.1.1.

\subsection{Performance-Feedback Iterative Optimization}

\begin{figure}[t]
    \centering
    \includegraphics[width=\textwidth]{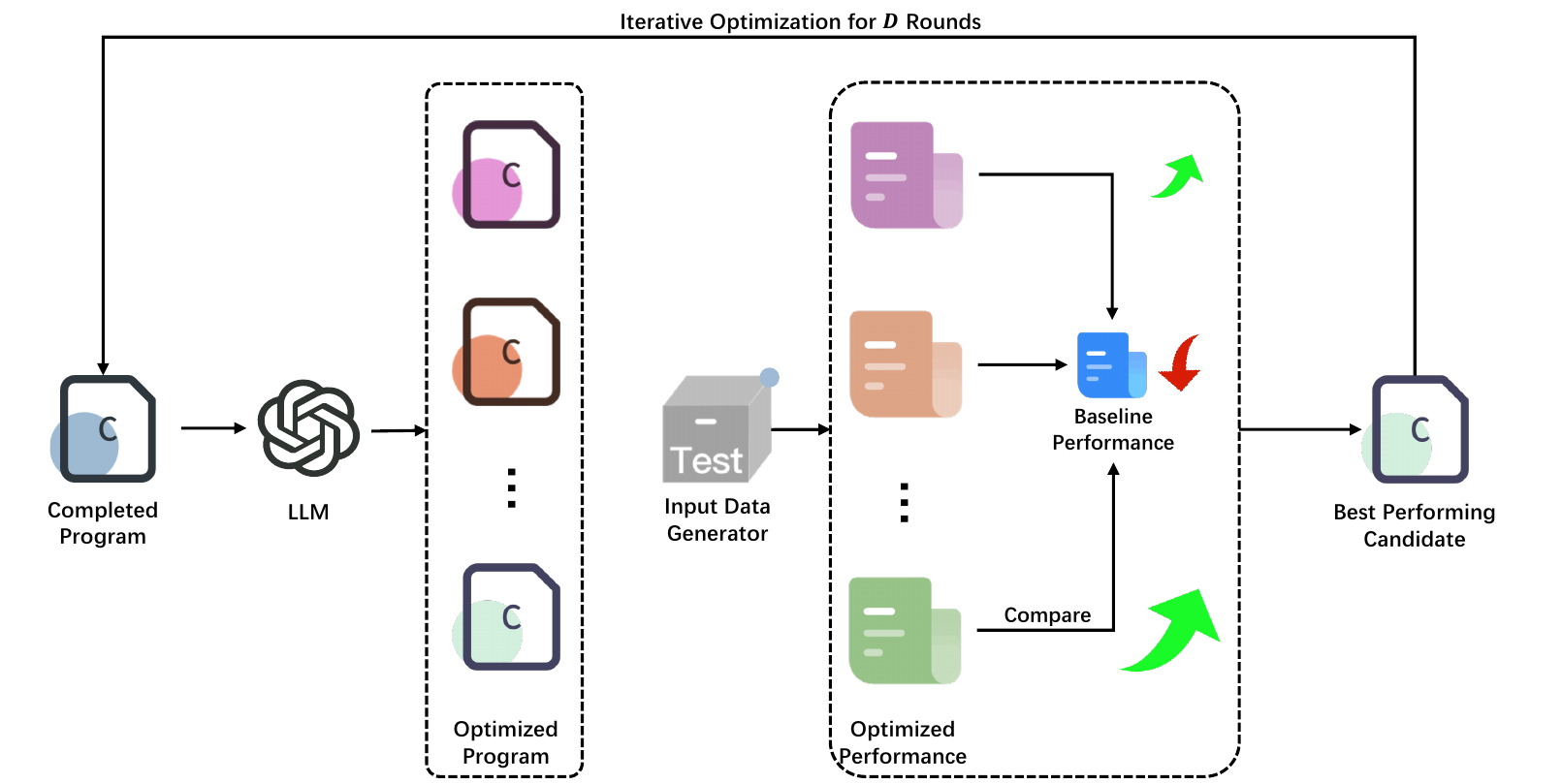}
    \caption{The framework of performance-feedback iterative optimization.}
    \label{fig:stage2}
\end{figure}

After the construction and evaluation of MEP, the framework enters Performance-Feedback Iterative Optimization. Each round generates multiple candidates; the best-performing candidate becomes the baseline for the next round, progressively converging to an improved solution. The framework of this stage is shown in Figure~\ref{fig:stage2}.

Let $K^{(0)}$ denote the baseline kernel version, and $K^{(d)}$ denote the baseline at the start of round $d$. Guided by profiler feedback (e.g., cache hit rate, occupancy), the LLM generates up to $N$ candidates $\{K^{(d,n)}\}_{n=1}^{N}$. Let $T_{\text{ker}}^{(d,n)}$ be the average kernel execution time of candidate $n$ in round $d$. We evaluate performance with repeated runs ($R$ times) and a trimmed mean to mitigate outliers due to system noise. Sort the $R$ measurements ascendingly as $T_{(1)}^{(d,n)}, \ldots, T_{(R)}^{(d,n)}$ and discard the lowest and highest $k$ values:

\begin{equation}
\overline{T_{\text{ker}}^{(d,n)}} = 
\frac{\displaystyle \sum_{i=k+1}^{R-k} T_{(i)}^{(d,n)}}{R - 2k},
\quad \text{with } R > 2k.
\end{equation}

Candidates must satisfy Functional Equivalence (FE) with respect to the baseline $K^{(d)}$, validated via output consistency checks. Define the feasible set

\begin{equation}
\mathcal{C}^{(d)} = \left\{ K^{(d,n)} \;\middle|\; 1 \le n \le N,\; \text{FE}\big(K^{(d,n)}, K^{(d)}\big) \right\}.
\end{equation}

We then select the next baseline by
\begin{equation}
n^\star = \arg\min_{1 \le n \le N,\; K^{(d,n)} \in \mathcal{C}^{(d)}} \;\overline{T_{\text{ker}}^{(d,n)}},
\qquad
K^{(d+1)} = K^{(d,n^\star)}.
\end{equation}

Iterations stop when $d = D$ or when the improvement falls below the preset threshold. Throughout, AER is invoked on compilation or runtime failures, or FE violations, to automatically repair candidates. In addition, effective optimization patterns (e.g., tiling choices, memory coalescing, synchronization restructuring) are summarized and injected as constraints or hints for subsequent rounds (Performance Pattern Inheritance), enabling the framework to reuse successful strategies and accelerate convergence.

\section{Experiment}

\subsection{Experimental Setup}

To comprehensively evaluate the effectiveness and portability of the proposed end-to-end LLM framework with performance feedback, we conduct systematic experiments on two hardware platforms, two benchmark suites, and several large-scale applications. This section presents the experimental environment, dataset configurations, parameter settings, and performance evaluation methods.

\subsubsection{Hardware and Software Environment}
\thinspace \par
\textbf{Platform 1 (NVIDIA GPU):} NVIDIA GeForce RTX 2080 Ti, CUDA 11.5, Ubuntu 22.04.

\textbf{Platform 2 (Haiguang DCU~\cite{ref30}):} An accelerator based on an AMD-licensed architecture, supporting the Heterogeneous-Compute Interface for Portability (HIP) programming model; HIP 24.04; CentOS 7.

\textbf{LLM Selection:} OpenAI o3, accessed via API and combined with our framework’s prompt and performance feedback strategy for iterative optimization.

\subsubsection{Datasets and Test Programs}
\thinspace \par
\textbf{PolyBench}~\cite{ref31}: We select 20 GPU kernel benchmarks covering matrix computations, sparse computations, and correlation calculations for testing on both NVIDIA and DCU platforms.

\textbf{AMD APP SDK}~\cite{ref32}: We select 12 representative GPU example programs, including image processing, signal transformation, and basic vector/matrix operations, tested only on the DCU platform.

\textbf{Large-Scale Supercomputing Applications:} Three HPC applications running on the Oriental Supercomputer (LAMMPS~\cite{ref33}, MISA-MD~\cite{ref34} and xblue) are characterized by complex engineering dependencies and large computational scale, and are tested only on the DCU platform. LAMMPS is an open-source molecular dynamics simulator widely used for materials modeling. MISA-MD is a molecular dynamics software. And xblue is a computational fluid dynamics code.

\subsubsection{Iterative Optimization Parameter Settings}

For PolyBench, we set $D=6$ iteration rounds and $N=3$ candidate codes per round, due to relatively small kernel sizes and limited optimization space.  
For AMD APP SDK and large-scale applications, we use $D=10$ and $N=5$ to fully explore optimization opportunities in more complex kernels.

Performance measurements: each candidate version is executed $R=30$ times, applying the trimmed mean method to remove the highest and lowest $k=3$ runs before averaging.

\subsubsection{Performance Evaluation Method}

We define Speedup as:
\[
\text{Speedup} = \frac{t_{\text{baseline}}}{t_{\text{optimized}}}
\]
where $t$ is the average execution time.

Three performance indicators are reported:
\begin{enumerate}
    \item \textbf{Standalone Speedup} – measured in the MEP.
    \item \textbf{Integrated Speedup} – measured after integrating the optimized kernel back into the original application.
    \item \textbf{Direct LLM Optimization Speedup} – measured by directly using the kernel generated from OpenAI o3 without performance-feedback loop, serving as a baseline comparison.
\end{enumerate}

Only optimizations passing Functional Equivalence verification are included in Speedup statistics.

\subsection{PolyBench Benchmark Results}

PolyBench is a standard GPU optimization benchmark suite derived from PolyBench, adapted for CUDA frameworks.  
Most kernels are small-scale with single computation paths, offering limited optimization space.  
We test on both NVIDIA RTX 2080 Ti (CUDA) and DCU platforms (HIP ported from CUDA).

Results are shown in Tables~\ref{tab:polybench-nvidia} and~\ref{tab:polybench-dcu}.

\begin{table}[!t]
\centering
\setlength{\tabcolsep}{12pt}
\caption{PolyBench speedup on NVIDIA platform}
\label{tab:polybench-nvidia}
\begin{tabular}{lccc}
\toprule
Name & Standalone & Integrated & Direct LLM Optimization \\
\midrule
2MM & 3.39 & 1.45 & 1.45 \\
3MM & 2.84 & 3.13 & 3.09 \\
ADI & 1.57 & 1.58 & 1.47 \\
ATAX & 2.19 & 1.06 & 1.05 \\
BICG & 7.14 & 7.22 & 1.00 \\
CORR & 14.64 & 14.27 & 5.42 \\
COVAR & 14.04 & 13.78 & 0.86 \\
GEMM & 2.75 & 3.48 & 3.63 \\
GEMVER & 7.10 & 7.23 & 3.59 \\
GESUMMV & 1.79 & 1.84 & 1.87 \\
GRAMSCHM & 6.55 & 5.89 & 1.73 \\
SYR2K & 3.70 & 3.74 & 1.02 \\
SYRK & 1.08 & 1.08 & 1.08 \\
\midrule
Average & 5.29 & 5.05 & 2.10 \\
\bottomrule
\end{tabular}
\end{table}

\begin{table}[!t]
\centering
\setlength{\tabcolsep}{12pt}
\caption{PolyBench speedup on DCU platform}
\label{tab:polybench-dcu}
\begin{tabular}{lccc}
\toprule
Name & Standalone & Integrated & Direct LLM Optimization \\
\midrule
2MM & 2.12 & 1.25 & 0.98 \\
3MM & 30.99 & 22.30 & 2.57 \\
BICG & 5.19 & 5.57 & 0.90 \\
CORR & 6.02 & 6.83 & 0.98 \\
COVAR & 11.87 & 9.00 & 0.84 \\
GEMM & 5.50 & 6.26 & 6.08 \\
GEMVER & 5.60 & 5.84 & 0.90 \\
GESUMMV & 3.77 & 4.05 & 1.08 \\
SYR2K & 2.63 & 10.60 & 4.56 \\
SYRK & 4.40 & 5.99 & 3.93 \\
\midrule
Average & 7.81 & 7.77 & 2.28 \\
\bottomrule
\end{tabular}
\end{table}

We observe consistent performance trends across both platforms, with average integrated speedups of $5.05\times$ (NVIDIA) and $7.77\times$ (DCU), both clearly outperforming Direct LLM Optimization ($2.10\times$ and $2.28\times$).  
Standalone test results closely match the integrated performance, validating the credibility of using MEP for optimization guidance.

\subsection{AMD APP SDK Benchmark Results}

AMD APP SDK provides diverse GPU sample programs and is optimized in terms of memory access and thread layouts.  
We test 12 representative cases on the DCU platform using HIP, with $D=10$, $N=5$.

\begin{table}[!t]
\centering
\setlength{\tabcolsep}{12pt}
\caption{AMD APP SDK speedup on DCU platform}
\label{tab:amd-sdk}
\begin{tabular}{lccc}
\toprule
Name & Standalone & Integrated & Direct LLM Optimization \\
\midrule
binomialoption & 1.26 & 1.25 & 0.99 \\
bitonicsort & 1.03 & 3.86 & 1.73 \\
dwthaar1d & 1.00 & 1.00 & Err \\
fastwalshtransform & 1.01 & 2.43 & 0.75 \\
matrixmultiplication & 1.15 & 1.19 & 1.10 \\
reduction & 3.38 & 1.94 & 1.26 \\
simpleconvolution & 1.03 & 1.04 & 0.85 \\
vectoradd & 1.05 & 1.47 & 1.21 \\
\midrule
Average & 1.36 & 1.77 & 1.13 \\
\bottomrule
\end{tabular}
\end{table}

The average integrated improvement is $1.77\times$, significantly higher than the $1.13\times$ from Direct LLM Optimization.  
Notably, kernels involving multi-thread synchronization and memory optimization (e.g., \textit{bitonicsort}, \textit{reduction}) show substantial acceleration.

\subsection{Large-Scale Applications Results}

We test three hotspot kernels of HPC applications running on the Oriental Supercomputer: LAMMPS (\texttt{k\_energy\_fast}), xblue (\texttt{gauss\_all\_seidel\_backfor}), and MISA-MD (\texttt{md\_nei\_itl\_wf\_atom\_soa}).  
Each has complex dependencies and larger computational scale than benchmarks.  
We use $D=10$, $N=5$ for these tests.

\begin{table}[!t]
\centering
\setlength{\tabcolsep}{12pt}
\caption{Large-scale HPC applications speedup on DCU platform}
\label{tab:large-app}
\begin{tabular}{lccc}
\toprule
Name & Standalone & Integrated & Direct LLM Optimization \\
\midrule
LAMMPS & 1.85 & 1.60 & 1.00 \\
MISA-MD & 1.01 & 1.01 & 0.98 \\
xblue & 2.48 & 1.13 & 1.00 \\
\midrule
Average & 1.78 & 1.25 & 1.00 \\
\bottomrule
\end{tabular}
\end{table}

The average integrated speedup is $1.25\times$, compared to nearly no improvement from Direct LLM Optimization.  
Standalone kernel results are closely aligned with integrated metrics, reinforcing the reliability of the MEP approach in complex settings.  
No cases showed expected gains in standalone tests but regressions upon integration, indicating strong robustness of the framework.

\section{Conclusion}
We proposed an end-to-end LLM-based GPU kernel optimization framework that builds Minimal Executable Programs to avoid costly full-application runs, integrating Automatic Error Repair and Performance Pattern Inheritance for efficient, cross-platform optimization. Experiments on benchmarks and real HPC kernels show clear gains over direct LLM optimization. While standalone MEP results generally predict integrated performance well, occasional larger gaps arise from inevitable differences between the MEP and the original execution environment, such as compiler, runtime, and dataflow discrepancies. Future work will focus on narrowing this gap by enhancing environment fidelity in MEP construction and adaptive integration strategies.

%
%
%
\bibliographystyle{unsrt} 
\bibliography{refs} 

@article{ref1,
  author  = {Chetlur, Sharan and Woolley, Cliff and Vandermersch, Philippe and Cohen, Jonathan and Tran, John and Catanzaro, Bryan and Shelhamer, Evan},
  title   = {cuDNN: Efficient Primitives for Deep Learning},
  journal = {arXiv preprint arXiv:1410.0759},
  year    = {2014},
  url     = {https://arxiv.org/abs/1410.0759}
}

@techreport{ref2,
  author = {Kerr, Andrew},
  title  = {CUTLASS: CUDA Templates for Linear Algebra Subroutines},
  institution = {NVIDIA},
  year   = {2019}
}

@article{ref3,
  author  = {Spector, Benjamin F. and Arora, Shubham and Singhal, Anirudh and others},
  title   = {ThunderKittens: Simple, Fast, and Adorable AI Kernels},
  journal = {arXiv preprint arXiv:2410.20399},
  year    = {2024},
  url     = {https://arxiv.org/abs/2410.20399}
}

@article{ref4,
  author  = {Dao, Tri},
  title   = {FlashAttention-2: Faster Attention with Better Parallelism and Work Partitioning},
  journal = {arXiv preprint arXiv:2307.08691},
  year    = {2023},
  url     = {https://arxiv.org/abs/2307.08691}
}

@inproceedings{ref5,
  author    = {Shah, J. and Bikshandi, G. and Zhang, Y. and Thakkar, V. and Ramani, P. and Dao, T.},
  title     = {FlashAttention-3: Fast and Accurate Attention with Asynchrony and Low-Precision},
  booktitle = {Advances in Neural Information Processing Systems},
  volume    = {37},
  pages     = {68658--68685},
  year      = {2024}
}

@inproceedings{ref6,
  author    = {Li, Chenshuang and Xu, Yao and Mahdipour Saravani, Siamak and Sadayappan, Ponnuswamy},
  title     = {Accelerated Auto-Optimization of GPU Kernels for Tensor Computations},
  booktitle = {Proceedings of the 38th ACM International Conference on Supercomputing (ICS '24)},
  pages     = {549--561},
  publisher = {ACM},
  year      = {2024},
  doi       = {10.1145/3650200.3656626}
}

@inproceedings{ref7,
  author    = {Wu, Minjia and Cheng, Xin and Liu, Shan and others},
  title     = {Mirage: A Multi-Level Superoptimizer for Tensor Programs},
  booktitle = {19th USENIX Symposium on Operating Systems Design and Implementation (OSDI '25)},
  pages     = {21--38},
  year      = {2025}
}

@inproceedings{ref8,
  author    = {Chen, Tianqi and Moreau, Thierry and Jiang, Ziheng and Zheng, Lianmin and Yan, Eddie and Shen, Haichen and others},
  title     = {TVM: An Automated End-to-End Optimizing Compiler for Deep Learning},
  booktitle = {13th USENIX Symposium on Operating Systems Design and Implementation (OSDI '18)},
  pages     = {578--594},
  year      = {2018}
}

@inproceedings{ref9,
  author    = {Zheng, Lianmin and Jia, Chengtao and Sun, Mingzhen and Wu, Zhen and Yu, Christopher H. and Haj-Ali, Ameer and others},
  title     = {Ansor: Generating High-Performance Tensor Programs for Deep Learning},
  booktitle = {14th USENIX Symposium on Operating Systems Design and Implementation (OSDI '20)},
  pages     = {863--879},
  year      = {2020}
}

@inproceedings{ref10,
  author    = {Ouyang, An and Guo, Shuo and Arora, Shubham and Zhang, Alex L. and Hu, Weijia and R{\'{e}}, Christopher and Mirhoseini, Azalia},
  title     = {KernelBench: Can LLMs Write Efficient GPU Kernels?},
  booktitle = {Proceedings of the 42nd International Conference on Machine Learning (ICML '25)},
  volume    = {267},
  series    = {PMLR},
  year      = {2025},
  note      = {arXiv preprint arXiv:2502.10517}
}

@article{ref11,
  author  = {Wen, Zhe and Zhang, Yue and Li, Zhongzhu and Liu, Zhiyuan and Xie, Lirong and Zhang, Tao},
  title   = {MultiKernelBench: A Multi-Platform Benchmark for Kernel Generation},
  journal = {arXiv preprint arXiv:2507.xxxx},
  year    = {2025}
}

@article{ref12,
  author  = {Chen, Mark and Tworek, Jerry and Jun, Heewoo and Yuan, Qiming and Pinto, Henrique Ponde de Oliveira and Kaplan, Jared and others},
  title   = {Evaluating Large Language Models Trained on Code},
  journal = {arXiv preprint arXiv:2107.03374},
  year    = {2021},
  url     = {https://arxiv.org/abs/2107.03374}
}

@article{ref13,
  author  = {Li, Yujia and Choi, David and Chung, Junyoung and Kushman, Nate and Schrittwieser, Julian and Leblond, Rémi and others},
  title   = {Competition-Level Code Generation with AlphaCode},
  journal = {Science},
  volume  = {378},
  number  = {6624},
  pages   = {1092--1097},
  year    = {2022}
}

@techreport{ref14,
  author = {Lange, Robert T. and Prasad, Ananya and Sun, Qian and Faldor, Max and Tang, Yichu and Ha, David},
  title  = {The AI CUDA Engineer: Agentic CUDA Kernel Discovery, Optimization and Composition},
  institution = {Sakana AI},
  year   = {2025}
}

@article{ref15,
  author  = {Chen, Weilin and Zhu, Jie and Fan, Qian and Ma, Yifan and Zou, Ang},
  title   = {CUDA-LLM: LLMs Can Write Efficient CUDA Kernels},
  journal = {arXiv preprint arXiv:2506.09092},
  year    = {2025}
}

@article{ref16,
  author  = {Andrews, Michael and Witteveen, Sander},
  title   = {GPU Kernel Scientist: An LLM-Driven Framework for Iterative Kernel Optimization},
  journal = {arXiv preprint arXiv:2506.20807},
  year    = {2025}
}

@article{ref17,
  author  = {Li, Jia and Li, Shuang and Gao, Zhi and Shi, Qiang and Li, Yifan and Wang, Zhen and others},
  title   = {TritonBench: Benchmarking Large Language Model Capabilities for Generating Triton Operators},
  journal = {arXiv preprint arXiv:2502.14752},
  year    = {2025}
}

@misc{ref18,
  author = {Zheng, Lianmin and Yin, Liang and Xie, Zhengda and Huang, Jiawei and Sun, Chen and Yu, Chen and others},
  title  = {Efficiently Programming Large Language Models Using SGLang},
  year   = {2023},
  note   = {Technical / preprint}
}

@article{ref19,
  author  = {Ragan-Kelley, Jonathan and Barnes, Connelly and Adams, Andrew and Paris, Sylvain and Durand, Frédo and Amarasinghe, Saman},
  title   = {Halide: A Language and Compiler for Optimizing Parallelism, Locality, and Recomputations in Image Processing Pipelines},
  journal = {ACM SIGPLAN Notices},
  volume  = {48},
  number  = {6},
  pages   = {519--530},
  year    = {2013}
}

@article{ref20,
  author  = {Lattner, Chris and Amini, Mehdi and Bondhugula, Uday and Cohen, Albert and Davis, Andy and Pienaar, Jacques and others},
  title   = {MLIR: A Compiler Infrastructure for the End of Moore’s Law},
  journal = {arXiv preprint arXiv:2002.11054},
  year    = {2020}
}

@article{ref21,
  author  = {Abadi, Martín and Agarwal, Ashish and Barham, Paul and Brevdo, Eugene and Chen, Zhifeng and Citro, Craig and others},
  title   = {TensorFlow: Large-Scale Machine Learning on Heterogeneous Distributed Systems},
  journal = {arXiv preprint arXiv:1603.04467},
  year    = {2016}
}

@misc{ref22,
  author = {Thakkar, Vaibhav and Ramani, Pradeep and Cecka, Curtis and Shivam, Abhishek and Lu, Hanlin and Yan, Eddie and others},
  title  = {CUTLASS (CUDA Templates for Linear Algebra Subroutines, Version 3.0.0)},
  year   = {2023},
  url    = {https://github.com/NVIDIA/cutlass/tree/v3.0.0}
}

@inproceedings{ref23,
  author    = {Taneja, Jay and Laird, Andrew and Yan, Cheng and Musuvathi, Madan and Lahiri, Shuvendu K.},
  title     = {LLM-Vectorizer: LLM-Based Verified Loop Vectorizer},
  booktitle = {Proceedings of the 23rd ACM/IEEE International Symposium on Code Generation and Optimization},
  pages     = {137--149},
  year      = {2025}
}

@misc{ref24,
  author = {Wei, Alexander and Suresh, Tanzim and Tan, Hao and Xu, Yifan and Singh, Gaurav and Wang, Kevin and Aiken, Alex},
  title  = {Improving Assembly Code Performance with Large Language Models via Reinforcement Learning},
  year   = {2025},
  url    = {https://arxiv.org/abs/2505.11480}
}

@inproceedings{ref25,
  author    = {Zhai, Yuan and Yang, Shuo and Pan, Kai and Zhang, Rui and Liu, Sheng and Liu, Chuan and others},
  title     = {Enabling Tensor Language Model to Assist in Generating High-Performance Tensor Programs for Deep Learning},
  booktitle = {18th USENIX Symposium on Operating Systems Design and Implementation (OSDI ’24)},
  pages     = {289--305},
  year      = {2024}
}

@article{ref26,
  author  = {Wei, Alexander and Cao, Jian and Li, Ruofan and Chen, Haotian and Zhang, Yizhou and Wang, Zhen and others},
  title   = {EquiBench: Benchmarking Code Reasoning Capabilities of Large Language Models via Equivalence Checking},
  journal = {arXiv preprint arXiv:2502.xxxx},
  year    = {2025}
}

@misc{ref27,
  author = {Agrawal, Lalit A. and Tan, Si and Soylu, Doga and Ziems, Caleb and Khare, Rohan and Opsahl-Ong, Kelly and others},
  title  = {GEPA: Reflective Prompt Evolution Can Outperform Reinforcement Learning},
  year   = {2025},
  url    = {https://arxiv.org/abs/2507.19457}
}

@misc{ref28,
  author = {Baronio, Claudio and Marsella, Pasquale and Pan, Bowen and Guo, Shuo and Alberti, Stefano},
  title  = {KEVIN: Multi-Turn RL for Generating CUDA Kernels},
  year   = {2025},
  url    = {https://arxiv.org/abs/2507.11948}
}

@misc{ref30,
  title = {Haiguang DCU official site},
  howpublished = {\url{https://www.hygon.cn/index}}
}

@misc{ref31,
  title = {PolyBench},
  howpublished = {\url{https://github.com/sgrauerg/polybenchGpu}}
}

@misc{ref32,
  title = {AMD APP SDK},
  howpublished = {\url{https://en.wikipedia.org/wiki/AMD_APP_SDK}}
}

@misc{ref33,
  title = {LAMMPS},
  howpublished = {\url{https://www.lammps.org/}}
}

@misc{ref34,
  title = {MISA-MD},
  howpublished = {\url{https://misa-md.github.io/MDoc/}}
}
\end{document}